\documentclass[prb,twocolumn,showpacs,preprintnumbers,amsmath,amssymb]{revtex4}
\usepackage{graphicx}
\usepackage{longtable}
\usepackage{dcolumn}
\usepackage{bm}

\begin{document}

\title{Effective Coulomb interaction in transition metals from
constrained random-phase approximation}

\author{Ersoy \c{S}a\c{s}{\i}o\u{g}lu}\email{e.sasioglu@fz-juelich.de}
\author{Christoph Friedrich}
\author{Stefan  Bl\"{u}gel}

\affiliation{Peter Gr\"{u}nberg Institut and Institute for
Advanced Simulation, Forschungszentrum J\"{u}lich and JARA,
52425 J\"{u}lich, Germany}

\date{\today}

\begin{abstract}
The effective on-site Coulomb interaction (Hubbard $U$) between
localized \textit{d} electrons in 3\textit{d}, 4\textit{d}, and
5\textit{d} transition metals is calculated employing a new
parameter-free realization of the constrained random-phase
approximation using Wannier functions within the full-potential
linearized augmented-plane-wave method. The $U$ values lie between
1.5 and 5.7 eV and depend on the crystal structure, spin
polarization, \textit{d} electron number, and \textit{d} orbital
filling. On the basis of the calculated $U$ parameters, we discuss
the strength of the electronic correlations and the instability of
the paramagnetic state towards the ferromagnetic one for
3\textit{d} metals.

\end{abstract}


\maketitle

Density functional theory (DFT) within the local-density
approximation (LDA) \cite{DFT_LDA} is a reliable method for
calculating ground-state properties of solids with weak electronic
correlations, i.e., for $U/W <1$, where $U$ is the effective
on-site Coulomb interaction between localized electrons (Hubbard
$U$) and $W$ is the bandwidth. However, LDA often fails to
describe systems with intermediate ($U/W \sim 1$) and strong ($U/W
>1$) electron correlations, such as transition-metal oxides,
rare earths, Kondo systems, etc. Properties of these materials are
usually calculated with phenomenological many-body Hamiltonians
such as the Hubbard\cite{Hubbard} or the Anderson impurity
model.\cite{Anderson} However, in these models the Coulomb and
also the one-particle hopping matrix elements are typically
empirical parameters that are determined such that the employed
model reproduces experimental results of interest.

For a long time, DFT-LDA and many-body model Hamiltonian methods
have been separate and complementary approaches. This has
drastically changed with the advent of the dynamical mean-field
theory (DMFT)\cite{Georges:96} that merged with LDA to a novel
computational method referred to as LDA+DMFT,\cite{DMFT} which
developed to a modern many-body approach for treating correlated
electron materials. In retrospect, the so-called LDA+$U$
method,\cite{LDA} an early attempt correcting the LDA functional
by introducing a simple mean-field-like Hubbard $U$ term for
localized \textit{d} or \textit{f} states, and today routinely
applied to a broad spectrum of systems can be regarded as its
static limit. Both LDA+$U$ and LDA+DMFT as well as other
approaches not mentioned here have in common that they rely on the
Hubbard $U$ as an additional parameter. Frequently the exact value
of $U$ is unknown which impedes the predictive power of these
approaches.

The problem of calculating the Hubbard $U$ parameter-free for
transition metals (TMs), i.e., from first principles, has been
addressed by several
authors.\cite{Kotani,Nakamura,SB_cLDA,cRPA_2,Miyake_2} A number of
different approaches have been proposed. Among them, constrained
local-density approximation (cLDA)\cite{SB_cLDA} is the most
popular one. However, cLDA is known to give unreasonably large $U$
values for late TMs due to difficulties in compensating the
self-screening error of the localized electrons.\cite{cRPA_2}
Furthermore, the frequency dependence of $U$ is unattainable. On
the other hand, the constrained random-phase approximation (cRPA),
though numerically much more demanding, does not suffer from these
difficulties. In contrast to cLDA, it also allows to access
individual Coulomb matrix elements, e.g., on-site, off-site,
intra-orbital, inter-orbital, and exchange.

The aim of this Rapid Communication is to present a systematic
study of the effective on-site Coulomb interaction (Hubbard
\textit{U}) between localized \textit{d} electrons in TMs
determined by means of first-principles calculations. Previous
cRPA studies of $U$ in TMs have focused only on the non-magnetic
(NM) state of the 3\textit{d} series and the results appeared to
be strongly dependent on the parameters used in the cRPA
schemes.\cite{cRPA_2,Miyake_2} In the present work we propose a
new simple parameter-free cRPA approach within the full-potential
linearized augmented-plane-wave (FLAPW) method using maximally
localized Wannier functions (MLWFs).\cite{Max_Wan} In addition to
NM and magnetic states of the 3\textit{d} series we consider
4\textit{d} and 5\textit{d} elements in the periodic table. We
discuss the strength of the electronic correlations and the
instability of the paramagnetic state towards the ferromagnetic
one for 3\textit{d} TMs on the  basis of the calculated Hubbard
$U$ parameters.

The basic idea of the cRPA is to define an effective interaction
$U$ between the localized \textit{d} electrons by restricting the
screening processes to those that are not explicitly treated in
the effective model Hamiltonian.\cite{cRPA_2} To this end, we
divide the full polarization matrix $P=P_d+P_r$, where $P_d$
includes only \textit{d}-\textit{d} transitions and $P_r$ is the
remainder. Then, the frequency-dependent effective Coulomb
interaction is given schematically by the matrix equation
$U(\omega) = [1-vP_r(\omega)]^{-1}v$, where $v$ is the bare
Coulomb interaction and $U(\omega)$ is related to the fully
screened interaction by
$\tilde{U}(\omega)=[1-U(\omega)P_d(\omega)]^{-1}U(\omega)$. The
static limit of the average diagonal matrix element of $U(\omega
\rightarrow 0)$ represented in a local basis can be regarded as
the Hubbard $U$ parameter.\cite{cRPA_2}

Although, cRPA is a general approach, its application to materials
with entangled bands is not straightforward. In these materials
the localized \textit{d} states that span the model subspace mix
with extended \textit{s} and \textit{p} states, and there is no
unique identification of the \textit{d}-\textit{d} transitions for
constructing $P_d$. Several procedures have been proposed in the
literature to overcome this problem. Aryasetiawan \textit{et
al.}\cite{cRPA_2} suggested to use an energy window or a range of
band indices to define the \textit{d} subspace. However, the
results depended strongly on the chosen window or band indices. An
alternative approach,\cite{Miyake_2} in which the hybridization of
the \textit{d} states was switched off, was not burdened by
additional parameters, but the $U$ values turned out to be
unphysically large for materials with strong
\textit{sp}-\textit{d} mixing, e.g., early TMs, and the unphysical
suppression of hybridization is unsatisfactory.

In the present work we propose a new parameter-free procedure
where $P_d$ is directly constructed from the definition of the
\textit{d} subspace. The latter is spanned by a set of
$\textrm{MLWFs}$ $w_{n\mathbf{R}}^{\alpha}(\mathbf{r})
=\frac{1}{N}\sum_{\mathbf{k},m}T_{\mathbf{R},mn}^{\alpha}(\mathbf{k})\varphi^{\alpha}_{\mathbf{k}m}(\mathbf{r})$,
where $N$ is the number of $\mathbf{k}$ points,
$T_{\mathbf{R},mn}^{\alpha}(\mathbf{k})$ is the unitary
transformation matrix,
$\varphi^{\alpha}_{\mathbf{k}m}(\mathbf{r})$ are single-particle
Bloch states of spin $\alpha$ and band index $m$, and $\mathbf{R}$
is the atomic position vector in the unit cell. We now define
$P_d$ as the polarization function that is generated by all
transitions that take place within the \textit{d} subspace. To
determine with what probability this applies to a given transition
between extended Bloch eigenstates $\varphi^{\alpha}_{\mathbf{k}m}
$ $\rightarrow$ $
\varphi^{\alpha}_{\mathbf{k}+\mathbf{q}m^{\prime}}$, which may be
mixtures of \textit{sp} and \textit{d} states, we multiply the
probability that the electron resides in the \textit{d} subspace
before the transition
$p^{\alpha}_{\mathbf{k}m}=\sum_{\mathbf{R},n}|T_{\mathbf{R},mn}^{\alpha}(\mathbf{k})|^2$
with the corresponding probability after the transition and obtain
$p^{\alpha}_{\mathbf{k}m}p^{\alpha}_{\mathbf{k}+\mathbf{q}m^{\prime}}$
as the probability for the transition itself. Thus, $P_d$ is
constructed from summing over all transitions in the Lehmann
representation multiplied with these probabilities. In this way,
the resulting effective interaction $U(\omega)$ only depends on
the  $\textrm{MLWFs}$ that span the \textit{d} subspace and is
basically independent of the used electronic structure method.

In a more formal approach, one can define $P_d$ as the density
correlation function
$P_d(\mathbf{r}t,\mathbf{r}^{\prime}t^{\prime}) = -i \langle
\Psi_0| \hat T [ \hat n_d(\mathbf{r}t), \hat
n_d(\mathbf{r}^{\prime}t^{\prime}) ] |\Psi_0 \rangle$ with the
Kohn-Sham determinant $\Psi_0$, the time-ordering operator $\hat
T$, and the Heisenberg density operator $\hat n(\mathbf{r}t)=\hat
n_d(\mathbf{r}t) + \hat n_r(\mathbf{r}t)$ decomposed according to
the \textit{d} subspace and the rest. Without time-dependent
external fields, $P_d$ only depends on the time difference
$t-t^{\prime}$. A Fourier transformation then yields the Lehmann
representation described above.

The ground-state calculations are carried out using the FLAPW
method as implemented in the \texttt{FLEUR} code \cite{Fleur}
within the LDA exchange-correlation potential.\cite{LSDA} The
MLWFs are constructed with the \texttt{Wannier90}
code.\cite{Wannier90,Fleur_Wannier90} We include six bands per TM
atom in the construction of the $\textrm{MLWFs}$,  i.e., the five
\textit{d} bands and the itinerant \textit{s} band. The matrix
elements of the effective Coulomb potential $U$ in the
$\textrm{MLWF}$ basis are given by $U_{\mathbf{R}n_1 n_3;n_4
n_2}^{\alpha\beta}(\omega) = \iint
w_{n_1\mathbf{R}}^{\alpha*}(\mathbf{r})w_{n_3\mathbf{R}}^{\alpha}(\mathbf{r})
U(\mathbf{r},\mathbf{r}^{\prime};\omega)
w_{n_4\mathbf{R}}^{\beta*}(\mathbf{r}^{\prime})
w_{n_2\mathbf{R}}^{\beta}(\mathbf{r}^{\prime})\:
d^3r\:d^3r^{\prime}$. The effective Coulomb potential
$U(\mathbf{r},\mathbf{r}^{\prime};\omega)$ itself is calculated
within the cRPA implemented in the \texttt{SPEX}
code.\cite{Friedrich} (for further technical details see
Ref.\,\onlinecite{Sasioglu}). We define the average on-site
diagonal (direct intra-orbital) and off-diagonal (exchange
inter-orbital) matrix elements of the effective Coulomb potential
as $U=\frac{1}{5}\sum_{n}^{(d)}U_{\mathbf{R}nn;nn}^{\alpha\beta}$
and $J=\frac{1}{20}\sum_{m,n (m\neq
n)}^{(d)}U_{\mathbf{R}mn;nm}^{\alpha\beta}$. The average
off-diagonal (direct inter-orbital) Coulomb matrix elements are
given by the relation $U^{\prime}=U-2J$. Although the matrix
elements of the effective Coulomb potential are formally
spin-dependent due to the spin dependence of the $\textrm{MLWFs}$,
we find that this dependence is negligible in practice.

\begin{figure}[t]
\begin{center}
\includegraphics[scale=0.53]{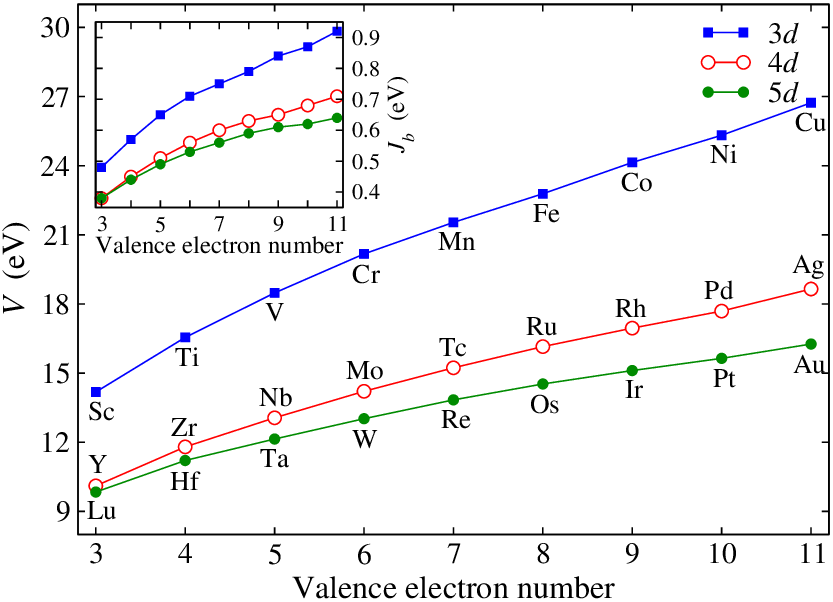}
\end{center}
\vspace*{-0.5cm} \caption{(Color online) Average bare on-site
direct Coulomb matrix elements between the \textit{d} orbitals for
TMs. In the inset we show the results for exchange Coulomb matrix
elements.} \label{bare_V}
\end{figure}

We start with the discussion of the \emph{un}screened (bare)
Coulomb interaction in the TMs. Figure\,\ref{bare_V} shows the
average bare on-site direct ($V$) Coulomb matrix elements for the
3\textit{d}, 4\textit{d}, and 5\textit{d} TM series in the NM
state. In the inset we show the results for exchange ($J_b$)
Coulomb matrix elements. Note that among the 3\textit{d} series
Fe, Co, and Ni are ferromagnetic (FM) while Cr orders
antiferromagnetically. Also Mn is FM in the bcc structure with
$a=2.91$ \AA. For these  elements we find that matrix elements of
the bare Coulomb potential for magnetic and NM states are nearly
identical. Within each series both $V$ and $J_b$ increase
monotonically with the \textit{d} electron number. This can be
explained by the fact that, as one moves from the left to the
right within one row of the periodic table, the nuclear charge
increases and causes the \textit{d}-wave functions to contract,
which gives rise to the observed trend for $V$ and $J_b$. On the
other hand, the localization of the \textit{d} electrons decreases
within one column of the periodic table from 3\textit{d} to
5\textit{d} elements. As a consequence, $V$ and $J_b$ decreases in
the same direction. This decrease is more pronounced for late
transition metals.

\begin{figure}[!ht]
\begin{center}
\includegraphics[scale=0.52]{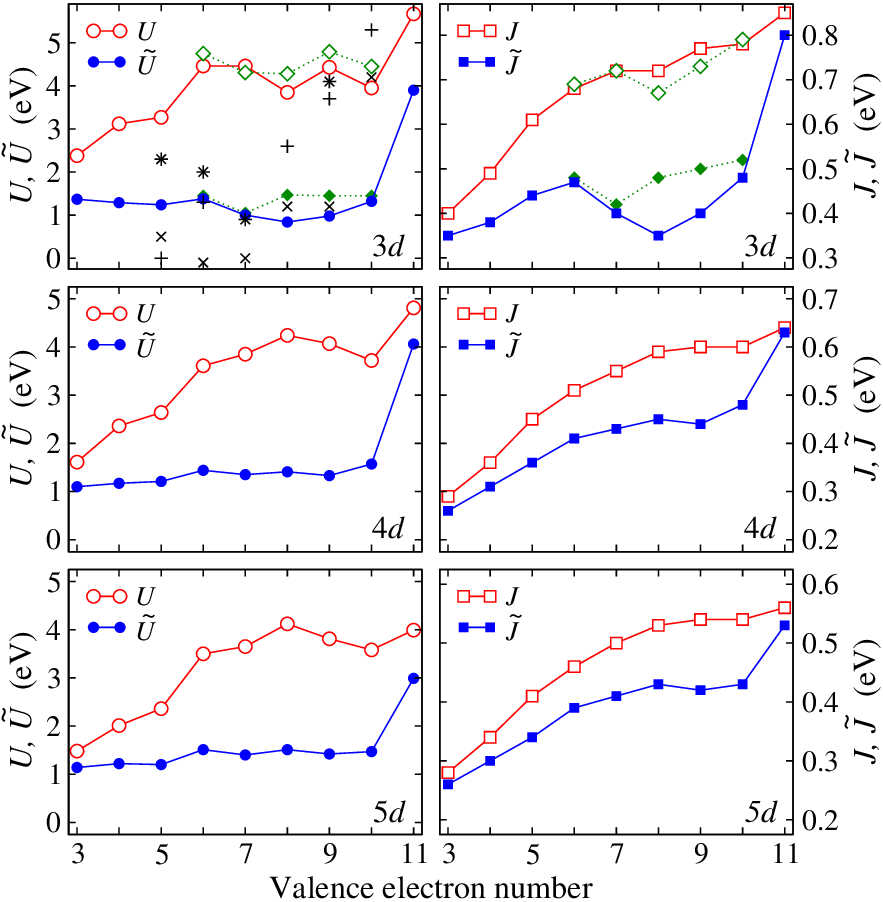}
\end{center}
\vspace*{-0.5cm} \caption{(Color online) Left panels: $U$ and
$\tilde{U}$ for the TM series. With open and filled diamonds we
show $U$ and $\tilde{U}$ for the magnetic state of 3\textit{d}
TMs. For comparison, experimental results from
Refs.\,\onlinecite{expt_1} (stars), \onlinecite{expt_2} (pluses),
and \onlinecite{expt_3} (crosses) as available are given. Right
panels: The same for $J$ and $\tilde{J}$.} \label{Hubbard_U}
\end{figure}

Efficient \textit{sp} screening in TMs significantly reduces the
bare Coulomb interaction $V$. Calculated Hubbard $U$ and $J$
parameters for the NM state of the TMs are presented in
Fig.\,\ref{Hubbard_U}. Results for the magnetic states of the
3\textit{d} elements are also included. For comparison, matrix
elements of the fully screened Coulomb interaction ($\tilde{U}$,
$\tilde{J}$) are given. As seen in Fig.\,\ref{Hubbard_U}, in
contrast to the bare direct Coulomb interaction $V$, the Hubbard
$U$ shows a non-monotonic behavior, i.e., it increases from the
early TMs and reaches a plateau-like behavior around half-filling,
whereas $\tilde{U}$ is almost constant across the TM series,
except for the elements with completely filled \textit{d} shells
like Cu. This behavior of the Hubbard $U$ reflects a substantial
contribution of the \textit{d}-\textit{d} transitions to the fully
screened Coulomb interaction $\tilde{U}$, especially around
half-filling. In difference to the bare Coulomb $V$, the Hubbard
$U$ parameter is very sensitive to the \textit{d} electron number
and \textit{d} orbital filing. In metals we are in the strong coupling 
limit, $v|P_r| \gg 1$, and thus $U \simeq -\frac{1}{P_r}$. Since $P_r$ depends 
mostly on the electronic structure of the screening electrons, this 
explains why isovalent TMs with the same crystal structure of 4\textit{d} 
and 5\textit{d} exhibit very similar $U$ values but different values of 
Mn, Fe, and Co, which have different crystal structures. 
Furthermore, by constrained NM and
proper spin-polarized treatments of the magnetic elements, we show
that spin polarization has a strong influence on $U$ and
$\tilde{U}$. The calculated $U$ values for Cr, Fe, Co, and Ni turn
out to be larger in the magnetic state than in the NM one, while
the situation is just the opposite for Mn, which was explained in
Ref.\,\onlinecite{Sasioglu} for the case of $\tilde{U}$ by the
different screening due to the available electrons at the  Fermi
energy. The same discussion holds also for partially screened $U$.
Our calculated  Hubbard $U$ parameters for the 3\textit{d} series
are in good agreement with recent cRPA studies of Miyake
\textit{et al.}\cite{Miyake_2} for late TMs as well as cLDA
calculations of Nakamura \textit{et al.}\cite{Nakamura} for early
TMs. Experimentally, the Hubbard $U$ parameters for 3\textit{d}
TMs are deduced from a combined use of Auger and x-ray
photoemission spectroscopy.\cite{expt_1,expt_2,expt_3} Results
from three different groups are included in Fig.\,\ref{Hubbard_U}
for comparison. As seen, the experimental $U$ parameters are
rather scattered. Our calculated $U$ values are in good agreement
with measurements of Kaurila \textit{et al.}\cite{expt_1} as well
as Yin \textit{et al.}\cite{expt_2} for late TMs. So far, we have
focused only on the effective intra-orbital direct Coulomb
interaction $U$. The same discussion holds also for the
inter-orbital direct and exchange Coulomb interaction $U^{\prime}$
and $J$, respectively. Note that in contrast to $U$ and
$U^{\prime}$, renormalization of the $J$ is rather small, i.e.,
$J$ is close to the atomic value $J_b$. However, the
\textit{d}-\textit{d} transitions substantially reduce
$\tilde{J}$, especially for late TMs. It should also be  noted
that while the bare $V$ has a long-range behavior, the $U$ shows
much faster damping. The calculated nearest-neighbor $U$ values
lie between 0.1 and 0.4 eV being maximal for TMs with half-filled
\textit{d} bands.

\begin{table}
\caption{Hubbard $U$ for $e_{\mathrm{g}}$ and $t_{2\mathrm{g}}$
orbitals (in eV) for bcc V, Nb, and Ta and fcc Ni, Pd, and Pt.}
\begin{ruledtabular}
\begin{tabular}{lcccccc}
        & V    & Nb   & Ta    & Ni  & Pd & Pt \\
         \hline
$U$($e_{\mathrm{g}}$)    & 3.47 & 2.78 & 2.58  & 4.04 & 3.76 & 3.63  \\
$U$($t_{2\mathrm{g}}$)   & 3.13 & 2.55 & 2.21  & 3.90 & 3.69 & 3.55  \\
\end{tabular}
\label{table}
\end{ruledtabular}
\end{table}

\begin{figure}[b]
\begin{center}
\includegraphics[scale=0.54]{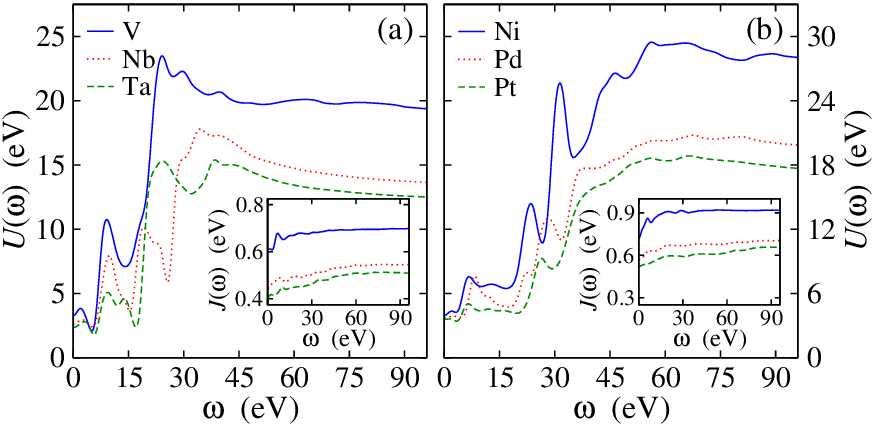}
\end{center}
\vspace*{-0.5cm} \caption{(Color online) (a) $U(\omega)$ for bcc
V, Nb, and Ta. The inset shows $J(\omega)$ for the same elements.
(b) The same as (a) for fcc Ni, Pd, and Pt.}
\label{Freq_dependence}
\end{figure}

In Table\,\ref{table} we present the orbital dependence of the
Hubbard $U$ for bcc V, Nb, and Ta and fcc Ni, Pd, and Pt.
Fig.\,\ref{Freq_dependence} shows the frequency dependence of $U$
and $J$ for the same elements. As can be seen, the crystal
structure has a sizable influence not only on the orbital
anisotropy of the Coulomb matrix elements but also on the
frequency dependence of $U$. For TMs with the bcc structure, the
effective interaction between \textit{d} electrons in
$e_{\mathrm{g}}$ orbitals are about 0.3 eV larger than
$t_{2\mathrm{g}}$ ones, whereas this difference is about 0.1 eV
for TMs having closed packed fcc and hcp (results not shown)
structures. For the former TMs, the $U(\omega)$ show strong
variations at low frequencies (see Fig.\,\ref{Freq_dependence}),
which suggests that the use of the static value $U(\omega=0)$ in
model Hamiltonians may be inappropriate. For the latter elements,
$U(\omega)$ shows a smoother behavior. At the plasmon frequency
(20-30 eV) the $U(\omega)$ increases rapidly. Above this frequency
the screening is not effective and $U(\omega)$ approaches the bare
value (compare Fig.\,\ref{bare_V}). In contrast to $U(\omega)$,
the exchange $J(\omega)$ is only weakly energy dependent and does
not show significant variations at the plasmon frequency.

\begin{figure}[t]
\begin{center}
\includegraphics[scale=0.54]{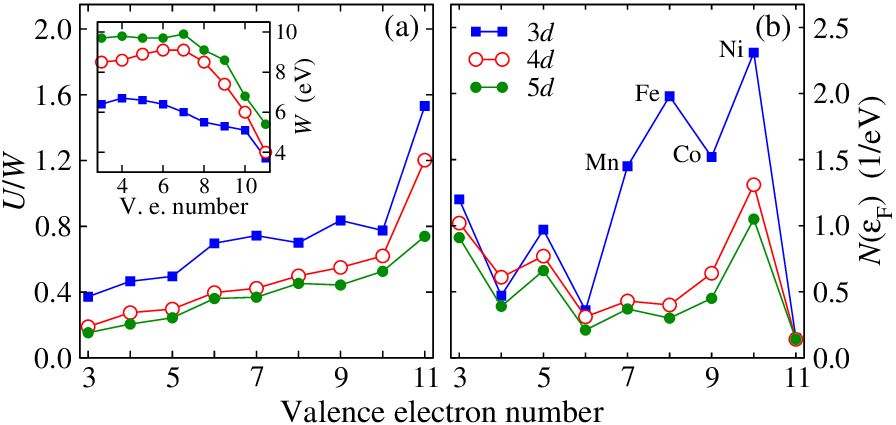}
\end{center}
\vspace*{-0.45cm} \caption{(Color online) (a) The ratio $U/W$ of
the effective Coulomb interaction $U$ and the \textit{d} bandwidth
$W$ for the TM series. The inset shows the \textit{d} bandwidths.
(b) Density of states $N(\epsilon_F)$ at the Fermi level for the
NM states.} \label{correlation}
\end{figure}

Finally, we  discuss the strength of the electronic correlations
and the instability of the paramagnetic state towards
ferromagnetism for the late 3\textit{d} TMs. In
Fig.\,\ref{correlation} (a) we show $U/W$ ratios for the TM
series, where the \textit{d} bandwidths $W$ are obtained from the
single-particle band structure. As seen, similar to the bare
Coulomb interaction $V$, the $U/W$ ratio increases from early to
late TMs and, as a result, the correlation strength increases.
Despite the similar $U$ parameters for isovalent TMs, the
3\textit{d} elements have larger $U/W$ ratios than 4\textit{d} and
5\textit{d} elements due to the much smaller bandwidths. For all
TMs, except Cu and Ag, we have $U/W<1$, which reveals weak
electronic correlations in these materials. Ferromagnetism of the
late TMs can be related to large $U/W$ ratios, but this condition
is not sufficient. In addition, the dimensionality and the crystal
structure, which dictates the shape of the density of states
(DOS), is crucial for the appearance of itinerant
ferromagnetism.\cite{Arita} In a mean-field treatment of itinerant
ferromagnetism the instability of the paramagnetic state is given
by the Stoner criterion $I N(\epsilon_F)>1$, where $I$ is the
Stoner parameter and $N(\epsilon_F)$ is the DOS at the Fermi level
in the NM state.  Using the Hartree-Fock solution of the
multi-orbital Hubbard model, Stollhoff \textit{et
al.}\cite{Stollhoff} proposed a relationship between the Stoner
parameter $I$ and  the Hubbard $U$ and $J$, which is given by
$I=(U+6J)/5$. These authors showed that electron correlation
reduces $I$ by roughly 40\%. Using the calculated $U$ and $J$
values we get $I=0.98$, 1.08, and 1.04 for Fe, Co, and Ni,
respectively, which is very close to the values 0.92 (Fe), 0.98
(Co), and 1.02 (Ni) obtained from linear-response
calculations.\cite{Mohn} Among the 3\textit{d} series, only Mn,
Fe, Co, and Ni satisfy the Stoner criterion due to the large DOS
at the Fermi level [see Fig.\,\ref{correlation} (b)], and the
paramagnetic state is unstable towards the formation of
ferromagnetism. The 4\textit{d} element  Pd is nearly
ferromagnetic. It shows strong spin fluctuations and exchange
enhancement.\cite{Paramagnons}

In conclusion, by employing a new parameter-free cRPA scheme we
have calculated the effective on-site Coulomb interaction (Hubbard
$U$) between localized \textit{d} electrons in TMs. We have shown
that the Hubbard $U$ depends on the crystal structure, spin
polarization, \textit{d} electron number, and \textit{d} orbital
filling, while it is insensitive to the \textit{d} character of
the elements. Most of the isovalent TMs assume similar \textit{U}
values. The obtained $U$ parameters for the 3\textit{d} TMs are in
good agreement with previous studies as well as available
experimental data and predict correctly the paramagnetic
instability towards the ferromagnetic state for the late
3\textit{d}s. The $U$ ($J$) values as calculated in the presented
approach increase considerably the predictive power of the LDA+$U$
and LDA+DMFT schemes applied to describe correlated electron
materials.

Fruitful discussions with F.\ Freimuth, A.\ Schindlmayr, T.\
Miyake, F.\ Aryasetiawan, and R.\ Sakuma are gratefully
acknowledged. This work has been supported in part by the DFG
through the Research Unit FOR-1346.

\end{document}